\documentclass{article}

\begin{document}

\bibliographystyle{plain}

{\bf \large WHAT IS THE MATHEMATICAL STRUCTURE OF QUANTUM SPACETIME?}

\bigskip

by: Louis Crane, math department, KSU

\bigskip

{\bf ABSTRACT:} {\it We survey indications from different branches of
Physics that the fine scale structure of spacetime is not adequately
described by a manifold. Based on the hints we accumulate, we propose
a new structure, which we call a quantum topos. In the process of
constructing a quantum topos for quantum gravity, we propose a new,
operational approach to the problem of the obervables in quantum
gravity, which leads to a new mathematical point of view on the state
sum models. }

\bigskip

{\bf I. INTRODUCTION}

\bigskip

The problem of finding the quantum theory of gravity has been a
fundamental challenge to theoretical Physics for almost a century
now. The perspective of this paper is that 
the mathematical description of spacetime as a smooth manifold 
inherited from classical
general relativity is inappropriate for the quantum theory and needs
to be replaced, and that this is the core of the problem. 

The discovery of general relativity by Einstein provides an
interesting model. Einstein was motivated by considering a rotating
wheel within special relativity \cite{St}. Since the circumference of the wheel
is along the direction of motion, it is contracted. The radius, on the
other hand, is not. One could simply explain this by saying that the
wheel is rotating. Einstein, however, was motivated by Mach's
principle, and believed that the only difference between a rotating
frame and one at rest was the relationship to the distant matter in
the universe. He therefore proposed that a gravitational field must
have the effect of changing the ratio of the radius of a circle to its
circumference. 

This one physical idea, motivated by a principle which in the end was
not exactly right, turned Einstein's attention to a branch of
Mathematics, differential geometry, which had the perfect structure to
understand gravity within relativity theory. Within differential
geometry, curvature, which can be measured by the deviation of the
ratio of the circumference of a circle to its radius from $2 \pi$ is
the central concept.

The adoption of a new mathematical description for spacetime was the
critical step in the process of
discovery of general relativity. The mathematical simplicity and
beauty of Riemannian geometry made the search for the equation of motion and its
physical interpretation tractable. In fact, Einstein made a wrong turn at one
point and formulated a theory he had to abandon. Nevertheless, the
mathematical structure guided him through. Einstein's equation was
almost the only possible equation of motion to write which is
generally covariant, i.e. well defined in the conceptual language of
differential geometry. Differential geometry has a symmetry principle,
namely general covariance, which was critical to expressing the
principle of relativity.

So, if we are correct that the
mathematical structure of a spacetime manifold needs to be replaced,
can we imitate Einstein, and find just the physical hint to identify
the new structure? Can we find some symmetry principle to guide our
search for a theory?

There are several reasons to hope this is possible. Relativists have
recently discovered a number of insights into the quantum theory. On the
other hand, mathematical approaches to space and geometry which do not
rely on an underlying point set are central to a number of areas of
pure Mathematics and have made many recent advances. In \cite{C1}, we made
a survey of these mathematical ideas, hopefully accessible to
Physicists, and suggested how they could be useful to quantum gravity.

The purpose of this paper is to attempt to construct a specific
candidate for the quantum structure of spacetime. In chapter 2,
we discuss a number of insights from relativity and quantum theory as to
the structure of spacetime, ending in a list of desiderata. In chapter
3, 
we propose a higher categorical
structure, called a quantum topos, which we will define, and whose
connections to the physical ideas we explore. In Chapter 4 we will
outline construction of a specific quantum topos intimately related to
general relativity and to the state sum models \cite{BC} for it as
well. 
This part of
the program is not finished, but we have an outline.

\bigskip

{\bf II. INSIGHTS FROM RELATIVITY AND QUANTUM MECHANICS INTO THE
  STRUCTURE OF SPACETIME. THE RELATIONAL SETTING}

\bigskip

{\bf A. Lessons from Quantum Mechanics}

\bigskip

There is a fundamental difference between the role of the variables
which describe the state of a system in classical and quantum
mechanics. In classical mechanics, the variables are regarded as all
having objective values, whether we measure them or not. In quantum
mechanics, on the contrary, quantities only take on values in the
course of a measurement process, which involves an external observer
system, and not all variables can attain values at once.

Realist approaches to quantum mechanics have been tried, but they have
uniformly failed. The operational approach to quantum mechanics is
central to our understanding of it.

What this tells us for quantum general relativity, is that the only
meaningful notion of position is apparent position, as viewed from an
observer outside the region of spacetime we treat as a quantum
process. An observer, treated as classical, which was located inside
the region, would cause the geometrodynamic process inside the region
to decohere. This motivates the following

\bigskip

{\large \bf DEFINITION} {\it The {\bf Relational setting} for a 
quantum theory of gravity is
a bounded region of
spacetime with a set of experimenters in its causal past who could
send probes into it, together with a set of observers in its future 
who could
detect the apparent positions of the probes.} 

\bigskip

The quantum theory of gravity must explain
correlations and regularities in the results of various experiments,
that is to say, the apparent positions of different probes as seen by
different observers in the relational setting.  

In order to interpret the theory, we need to assume that the
particular quantum state of our spacetime is reproducible, so that we
can think of ensembles of experiments and study relative
probabilities.

A very important point in the interpretation of quantum mechanics is
that the act of measurement disturbs the system. In the context of
quantum gravity, that means that it is not possible to think of
experiments involving idealised probes that do not react on the
spacetime geometry. Theory must explain only real experiments, in
which forming a black hole is an extreme possibility.

An important aspect of the interpretation of quantum mechanics is that
``world elements'', or propositions about a system which are either
true or false, do not form a distributive lattice ( see appendix
B). This contrasts to the subsets (open or measurable) of a manifold.

We are led to consider the possibility that the relational setting
will lead to a nondistributive lattice of observable subregions in
quantum spacetime.

\bigskip

{\bf B. Relative observation and curvature in general relativity}

\bigskip
The fundamental equation of general relativity relates the curvature
of the spacetime manifold to the distribution of matter. The curvature
of a manifold can be understood as the rate at which nearby geodesics
deviate. Thus, it is possible to understand classical general relativity as a
set of laws which determine where the same event will appear to be to
different observers. Naively, one might think that an observer would
interpret 
the apparent
positions of events as located in the causal past of a copy of
Minkowski space whose zero point coincides with the observer. In fact,
however, the curvature of the spacetime can cause a shear in the
congruence of null geodesics radiating from an event. This means that
apparent positions of past events occupy points in a circle bundle
over the past light cone of the observer. This fact, which seems to
have gone unremarked, plays a very important role in the quantum
theory of apparent position, as we shall discuss below. 

Neglecting this effect, by
observing correlations between where particular events appear to be to
several observers, an experimenter could identify the causal pasts of
different observers and reproduce the spacetime manifold, or at any
rate the part of it the observers could see. This is an operational
version of the standard definition of a manifold in terms of charts
and transition functions.  The observers in such a process need to
have two eyes, or a wide field of vision, in order to detect parallax
information. 

Thus, the data of classical general relativity would appear in the
relational setting as correlations between where different observers
see the same probe. It would be an interesting and useful exercise to
formulate general relativity explicitly in such a form.  

Now let us try to imagine how the correlations between apparent
positions for different observers might change as we go from classical
to quantum general relativity. Imagine for a moment a quantum state of
the metric for the region which was a superposition of two classical
spacetime metrics. A region in one metric would not appear to be
consistently within any region of the other metric of the same size to
all observers, if the region were small compared to the overall
curvature.  On the other hand, it would appear to be inside a larger
region in the second metric to any observer. This leads directly to a
failure of the distributive law for observable regions, consistent
with the behavior of ``world elements'' in quantum mechanics (see
Appendix B).

\bigskip

{\bf C. The Planck scale}

\bigskip

The oldest indication from Physics that the continuum of classical
Mathematics might not apply to physical spacetime is the existence of
the Planck scale. This length, which is far too small to be 
practically observable, is the result of the
interaction of general relativity with quantum mechanics.

To recapitulate for the mathematical reader, quantum mechanics tells
us that the uncertainties of position and momentum of any body are
inversely related. In order to confine it to a small space, it must
have a high probability of having a large momentum, and thus a large
energy.

General relativity, on the other hand, tells us that any body deforms
the causal structure of the spacetime around it. This is closely
related to the red shift near any massive body. If the energy density
in a small region is sufficiently high, it creates a black hole around
itself, so that no information from it can ever escape to the distant
universe. This can be thought of as an infinite red shift.

The combination of these two effects means that no length less than a
certain scale, called the Planck length or $l_p$ can ever be observed,
because any probe concentrated in so small a region would contain so
much energy that it disappeared into a black hole.
This distance is approximately $10^{-33}$ centimeter, much too small
to be detected directly.

In a classical theory this plays a minor role. However if we insist on
an operational approach to quantum geometry, it points to a very
profound departure from the classical continuum point set.

In the context of the relational setting, this places serious limits
on the amount of information about the geometry which can be
observed. 
More recent developments within
classical and semiclassical general relativity extend this quite
considerably, as we shall discuss.

\bigskip

{\bf D. Lessons from quantum field theory. Dimensional regularization.} 

\bigskip

The development of quantum field theory provides similar suggestions
that the classical spacetime continuum needs replacing. In the first
place, there is the problem of the ultraviolet divergences. Terms in
the Feynman perturbation series can be indexed by Feynman diagrams. To
compute the actual contribution of a given diagram to an amplitude, we
must take the integral over all ways to embed the diagram into
spacetime, inserting propagators related to the spacetime geometry on
the edges. These integrals are not in general finite. The important
divergences are concentrated in the region of the multiple integral
where one or more loops shrinks to zero length. 

This problem is solved by a complicated subtraction procedure called
renormalization. The solution has no real theoretical motivation;
different renormalizations can give different answers to the same
problem.

If we try to treat gravity as a quantum field theory, the subtraction
procedure breaks down. It is a ``nonrenormalizable'' theory.

The fact that all theoretical progress in particle Physics for the
last half century has depended on finding tricks that have no
conceptual justification has colored the entire field. I think it is
not duely recognized that such successes must be taken as provisional.

Renormalization generally begins by cutting off the distance scales at
some minimal length, for example by doing the theory on a discrete lattice.

This seems to suggest that the short distance structure of spacetime
is very different from a classical continuum.

The most successful theories in particle physics are the nonabelian
gauge theories. The development began with the discovery by t'Hooft
and Veltman \cite{TV} that gauge theories can be renormalized via dimensional
regularization. Before the discovery of dimensional regularization,
nonabelian gauge theory was believed to be nonrenormalizable.

In dimensional regularization, spacetime is treated as if it had
$4-\epsilon $ dimensions. This is accomplished by analytically
continuing the evaluations of integrals as a function of the spacetime
dimension. There is no mathematical understanding of the meaning of
this procedure, despite its central role in our understanding of
nature.

We need to be wary of the common misunderstanding that dimensional
regularization (or renormalisation in general) is a computational
device. The quantities to be calculated are mathematically undefined
until the renormalisation procedure is chosen, so it forms part of the
hypothetical structure of the theory. As such the meaning of
dimensional regularisation is not understood. If it is not a
hypothesis about the fine structure of spacetime, I have no idea what
it could be.

One could try to interpret this procedure by saying that at very short
scales, the number of observable regions of a given size was different from what
one would expect by counting subsets of a continuum.

\bigskip

{\bf E. Black hole thermodynamics and holography}

\bigskip

A series of discoveries about the classical and semiclassical behavior
of black holes has led to the idea that a black hole should be
regarded as having a finite entropy, equal to the area of its boundary
measured in Planck units.

\bigskip

$E=A/(l_p)^2$

\bigskip
This is supported by results which describe collisions of black holes,
matter falling into black holes, and most strikingly, by black hole
radiation. 

Since no information from the interior of a black hole can escape, it
is natural to interpret this as due to the state of the quantum fields
in the region of the skin of the black hole. A naive counting of all
modes in the quantum fields would give an infinite dimensional Hilbert
space, and therefore an infinite entropy. In order to give a
statistical explanation of the entropy ( = logarithm of the number of
microstates) it is necessary to assume the quantum field is cut off at
the energy corresponding to the Planck length.

Now a black hole is a global phenomenon, nothing remarkable happens
near its boundary locally. So if we assume that QFT is cut off there,
it should be cut off everywhere. This is further indication of the
nonapplicability of the classical continuum.

The ideas about black hole thermodynamics have implications for
arbitrary bounded regions of spacetime. In a body of work which goes
by the name of holography, it has been shown that any bounded region
of a suitably causal spacetime can only pass to its environment
information from a finite dimensional Hilbert space, with dimension
given by the entropy of a black hole which would replace the region,
i.e. its area in Planck units (whence the analogy to a hologram.)

This has immediate implications for the relational setting we have
been proposing. Only finitely many probes can be admitted without
turning the region into a black hole, which emits only thermal
noise. This means the description of the relational setting will be
cut off, and contain only finite information for any given region.

\bigskip

{\bf F. State sum models}

\bigskip

Recently a model \cite{BC} (more precisely, a closely connected set of models)
has been proposed for quantum general relativity in four dimensions. 
These models are associated with a four dimensional simplicial
complex, which can be thought of as a triangulation of a spacetime
manifold, but does not need to be.

In order to construct the models, we need to make use of the unitary
representations of the Lorentz group, as defined by Gelfand and 
Harish-Chandra. (In some variants, we use the representations of the 
Quantum
Lorentz Algebra instead). These representations denoted $ R( k, \rho
)$, are indexed by one real and one integer parameter, which can be
thought of as forming a complex spin $ 1/2 k+ i \rho $ , since the
lorentz algebra is a complexification of the ordinary 3D rotation
algebra.

In constructing the model, we used only intertwining operators between
representations; in other words, everything we did was manifestly
Lorentz invariant. This has the possibility of playing the role for
quantum gravity that general covariance did for classical general
relativity.

The model is constructed by assigning irreducible representations to
the 2-faces of a triangulation of the spacetime, putting in a special
intertwiner across each tetrahedron, tracing around each four simplex,
multiplying over four simplices, then summing over all labelings
\cite{BC}.

The representations are quantizations of the directed area elements,
or bivectors, which describe the areas and orientations of the faces. 

In order to be a quantization of the bivectors on triangles only, the
irreducible representations must satisfy a constraint; they must be
balanced. In the above notation, k=0.  This causes the resulting state 
sum to be nontopological,
because the balanced representations do not close under tensor
product. We shall see below that this recurs as a central question in
the quantization of apparent positions as well.

A significant number of the fundamental geometrical variables in
Minkowski spacetime, such as areas volumes and dihedral angles, have
natural quantizations as operators on these representations and their
tensor products. The geometry of a composite system (several faces,
for example), can be quantized on the tensor product of the
quantizations of the pieces.

In studying the state sum models we discovered that not all the
operators corresponding to geometric variables commute with one
another. For example, the operators giving the two shape parameters of
a tetrahedron, which can be thought of as two adjacent dihedral angles,
do not commute.

The state sum models give us several suggestions about how regions in
quantum spacetime should be described:

\bigskip

1. Unitary representations of the Lorentz algebra should appear as
building blocks, and all maps between them should intertwine Lorentz
symmetry. 

\bigskip

2. Regions should be modelled as tensor categories, in which parts are
tensored to make a whole. This is also suggested by standard quantum
mechanics as applied to compound systems, such as multiple spins.

\bigskip

3. The state sum models themselves should appear as approximations to
the full quantum theory of gravity. 

\bigskip

4. The geometrical variables will not all commute.

\bigskip

{\bf SUMMARY}

\bigskip

We have made the observation that four important areas of fundamental
Physics, renormalizable quantum field theory, gauge theory, 
classical general relativity in combination with quantum mechanics,
and the theory of semiclassical black holes,
all point to the conclusion that the classical continuum model of
spacetime must be abandoned. 

Furthermore the clues which lead us to this conclusion are in every
case central to the field of Physics from which they come. 
They are steps without clear justification, without which the theory
is unable to make contact with experiment.

It is a bit mysterious why these ideas have not been connected more
prominently, or taken as indicating an important direction. Perhaps
the chasm which has opened between Mathematics and Physics in our
educational system has something to do with it.

We have now found a number of physical indications of what the
quantum theory of spacetime should look like. We collect these for
future reference, connected to some mathematical ideas they seem to
suggest.

\bigskip

{\bf \large list of desiderata for quantum spacetime}

\bigskip

{\bf \it $ \alpha$ : minimum length scale- no point set continuum}

\bigskip

{\bf \it$ \beta$ : relational geometry-sheaves over a site of observers}

\bigskip

{\bf \it $\gamma$ : finite information transfer- red shift as filter}

\bigskip

{\bf \it $\delta$ : non-commutative variables-operational interpretation}

\bigskip

{\bf \it $\epsilon$ : Lorentz invariance-Gelfand representations as building
  blocks, and Lorentz covariant functions between them.}

\bigskip

{\it $\zeta$ : The density of regions of a given size should follow a
  different power law from that naively predicted by a spacetime
  manifold}

\bigskip

{\it $\iota$ : The regions of spacetime should not obey the distributive
  law.}

\bigskip

In the following sections, we shall show how a mathematical
construction called a quantum topos naturally accommodates these
criteria.

\bigskip

{\bf III. QUANTALOIDS AND THEIR SHEAVES }

\bigskip
{\bf A. Quantaloids for Physicists.}

\bigskip

Before giving a mathematical introduction to the type of mathematical
structure I am proposing for quantum spacetime, I want to explain
what they are and how they contain what the theory should contain 
in simple physical terms.

Since we cannot detect points, the geometry we can observe consists of
regions. Regions can contain one another, we can form unions and
intersections of them, and these operations satisfy obvious algebraic
laws.

Such structures are called lattices (no relation of the kind quantum
field theory is sometimes done on) see appendix A. We can construct a lattice by
starting with a set of points, and specifying certain subsets to be
members of the lattice.  Not every lattice can be constructed in this
way, however, and it is not necessary to assume that the physically
observable regions have a point set interpretation.

So we could begin by replacing the spacetime manifold with a
lattice. This would be a theory somewhat similar to causal sets, in
that the discrete point set in the causal set picture has a partial
order on it.

However, this would be an ``absolute'' spacetime, not a relational
one.  So a better approximation would be to describe the structure of
a spacetime as a lattice of observable regions for each
observer. There would then need to be some consistency relations
between what different observers see, which would be expressed by
``lattice maps''. 

So we need a structure of many lattices with maps of lattices between
them to construct a relational spacetime. This is what is called a
quantaloid. The ``quant'' in the name refers to quantum mechanics.

It is interesting that the history of the theory of quantaloids 
is a convergence of
two very different area of Mathematics. One is a natural but abstract
direction in higher category theory, the other is an attempt to
interpret quantum mechanical systems as noncommutative geometries.

\bigskip

{\bf B. Quantaloids, definitions and examples.}

\bigskip

{\bf DEFINITION}: {\it A quantaloid is a category enriched in the category
of sup-complete lattices.}

\bigskip

See appendix A for the definition and basic properties of a sup
complete lattice.

Let us unpack this for the non-categorical reader. There is a set of
objects, called the objects of the quantaloid. Between any two objects
A and B
there is a sup complete lattice called Hom (A,B), and for any three
objects A, B, C, of the quantaloid there is a tensor morphism of sup
complete lattices
\bigskip

 $C_{A,B.C} : Hom (A,B) \times Hom (B,C) \rightarrow Hom
(A, C) $  
\bigskip

called composition.

In this definition, $ \times$ denotes the tensor product of sup
complete lattices.

The operation of composition must distribute over arbitrary sups,
i.e.:

\bigskip

$a \circ (\vee b_i) = \vee (a  \circ b_i ) $;

\bigskip

this is a consequence of the definition of an enriched category which
we shall omit here.

\bigskip

{\bf Examples of quantaloids}

\bigskip

1. The free quantaloid on a category.

\bigskip

Given any category C we can define a quantaloid P(C) whose objects are the
objects of C. If A and B are objects of C, then the lattice
$Hom_{P(C)}(A,B) $ is the lattice of subsets of $Hom_C (A,B)$. The
composition is defined as 

\bigskip

$U \circ V =\{ uv \mid u \in U; v\in V \} $. 

\bigskip

So we can see that any mathematical structure has a quantaloid
version, where we focus on subsets instead of elements.

\bigskip

2. Quantales

\bigskip

{\bf DEFINITION:} {\it a quantaloid with one object is a quantale.}

\bigskip

Unpacking, this means that we have a sup complete lattice (the maps
from the object to itself) with a multiplication which distributes
over joins.  Any semigroup gives us a quantale, namely the lattice of
its subsets with 

\bigskip

$A \circ B = \{ ab: a \in A, b \in B \}$.

\bigskip

Algebraic structures richer than semigroups have quantalic versions
which reflect their additional operations. For any vector space we can
construct the quantale whose objects are subspaces and whose product
is given by

\bigskip

$ A \circ B = \{  \Sigma x_i a_i b_i : x_i  \in C, a_i  \in A, b_i
\in B  \} $ ;

\bigskip

where C is the complex numbers.

\bigskip

This is a subquantale of the quantale of all subsets of the 
vector space
under + , and can be constructed by operating on that quantale 
with the
operation of taking the linear closure. This is an example of a
quantalic nucleus, as we shall explain below.

Rings, algebras, and especially C*-algebras provide physically
interesting examples of quantales. The subsets of a ring form a
quantale under multiplication. The additive subgroups form a
subquantale, given as the image of the first under the operation of
additive closure, another quantalic nucleus. Left, right and 2-sided
ideals also form quantales. Similarly, the subspaces and ideals of an
algebra form quantales, where we have closed under the operation of
forming the linear span. 

An especially interesting example for Physics is the quantale of
closed right
ideals $ Q_{R^-} (A)$, where A is a $C^*$-algebra. The objects of this
quantale are the closed right ideals of A, and 

\bigskip

$B \circ D = CL \{ \Sigma x_i b_i d_i c_i \in C, d_i \in D, b_i \in B \} $ ;

\bigskip
where CL means topological closure, $x_i$ are complex numbers, and B
and D are closed right ideals of A.

In the case of a commutative $C^*$ -algebra, this reduces to the spectrum
of the algebra, considered as a locale. Every quantum mechanical
system has a quantale associated with its $C^*$-algebra of observables,
which can be thought of as a non-commutative geometric structure. 

\bigskip

3. Rel(C)

\bigskip

To any category C we can associate the quantaloid Rel(C) of
relations. The category of relations of the category of sets has sets
for objects and the lattice of subsets of $A \times B$ as
Hom(A,B). This is equivalent to substituting relations for functions.

To generalize this to an arbitrary category C we first construct the
category of spans of C. The objects are just the objects of C, while a
span between two objects A and B of C is a diagram:

\bigskip

$ B \leftarrow D \rightarrow B $

\bigskip

where D is any object of C.

Intuitively, we can imagine the graph of this diagram as a ``subset `` 
of $B \times
C$. Now to solve the problem that different spans might have the same
graph, we make a rather technical definition 

\bigskip

{\bf Definition:} a crible is a set of spans which whenever it
contains

\bigskip

$ B \leftarrow D \rightarrow B $

\bigskip

also contains the span generated by composing with any map $ E
\rightarrow D$. 
\bigskip

Cribles correspond to subsets of the cartesian product in the category
of sets. In a general category, the cribles between two objects form a
sup complete lattice,
and in general form a quantaloid.

\bigskip

{\bf C. Quantaloidal nuclei and Grothendieck topologies.}

\bigskip

We have a very general method of forming new quantaloids from old ones
by forming a quotient quantaloid. This involves the concept of a
quantaloidal nucleus.

\bigskip

{\bf Definition :} Let Q be a quantaloid. A quantaloidal nucleus on Q is
an assignment of a map $j_{a,b}$ for each pair of objects a,b of Q
such that:

\bigskip

1. $f \leq j(f)$

\bigskip

2. $ j^2=j$

\bigskip
 
3. $j_{a,b}(f) \circ j_{b,c}(g) \leq j_{a,c} (f \circ g)$

\bigskip
Now it is a well known theorem that quantalic nuclei correspond 1-1 to
quotient quantaloids. The image of any j satisfying 1-3 is again a
quantaloid, and any quotient quantaloid is so obtained. Furthermore,
any quantaloid is a quotient of a free quantaloid (example 1).

Quantaloidal nuclei are a generalisation of closure operators. The
quantaloidal nucleus which assigns to any subset of a vector space
its linear span gives the quantale of linear subspaces as a quotient
of the quantale of subsets. Similarly, the quantale of right ideals of
an algebra can be obtained from the quantale of its linear subspaces
by letting j give the ideal spanned by the subspace.

If we take any category C, the quantaloid Rel(C) has the property that
Grothendieck topologies on the site of C correspond to quantalic
nuclei on Rel(C) which respect intersection. In fact quantaloids were
invented, although not so named in \cite{P} precisely bacause they are
an easier to work with approach to Grothendieck topologies.

Any topos can be constructed as sheaves over  some quantaloid. The
quantaloids that so appear are like many-object versions of locales,
they have commutative multiplication and distributive lattice structure.
So we see that the suggestion below that quantum gravity lives in the
sheaves over a
quantaloid is a direct generalisation of the suggestion that it lives
in a topos, and can be thought of as a noncommutative version, or a
quantization of the older hypothesis.

\bigskip

{\bf D. Quantum topoi and geometry.}
\bigskip

Now there is a natural notion of presheaves over a quantaloid. It
consists of a set over each object of the quantaloid, and a member of
the lattice $ L_{x,y}  \in Hom(A,B)$ for each pair of one element x in 
the set over A
and one y in the set over B; satisfying the consistency relation:

\bigskip

$ L_{x, y} \times L_{y,z} \leq L_{x,z} $

\bigskip

which the categorically minded will recognize as the definition of a
weak or lax functor.

Sheaves are defined by the usual glueing property on all open covers.

In the quantaloidal picture there is a unique topology associated to
the quantaloid. We would pass to the analog of a new Grothendieck
topology 
(appendix
A) by passing
to a quotient  quantaloid.

Furthermore, the category of sheaves over a quantaloid is equivalent to the
category of presheaves \cite{G} , so the theory of quantaloidal nuclei has
fully 
absorbed the subtleties of Grothendieck topology.
 
Let us make the following 

\bigskip

{\bf Definition:} the category of sheaves over a quantaloid is a
quantum topos.

\bigskip

The sheaves over a quantaloid themselves
form a quantaloid.

If the quantaloid is a locale, we reproduce the ordinary definition of a
localic topos. 

To repeat,  our suggested definition of a quantum topos extends the
concept of a topos from commutative to noncommutative geometry.

Now quantum topoi can be viewed in many ways. They are equivalent to
categories enriched over the base quantaloid, and to variable sets, or
sets with relative equality relations.

One of the interesting ways of thinking of quantum topoi is that they
represent a kind of geometry. The example due to Lawvere \cite{L} illustrates
this. We define the locale $ R^{ + \leq }$ as the lattice of sets of the
form $ \{ x: 0 \leq x \leq a  \} $ for some positive a. 

Presheaves over this locale are exactly metric spaces. The lax
condition mentioned above gives the triangle inequality. Sheaves are
cauchy complete metric spaces, and the equivalence of the two
categories comes about via cauchy extension of functions.

If we think of these presheaves as sets with variable equality we get
the picture that two points in a metric space distance r
apart are equal up to stage r. This is an interesting precursor to a
relational geometry in which observers receive only finite
information.

Now the quantale in the above example is a locale, so commutative and
distributive. It is not surprizing the geometry it generates is
classical. Passing to a suitable noncommutative quantale is a natural
road to a quantum geometry.

Another important aspect of topos theory is that by doing Mathematics
in a topos we adopt its logical structure, which in the case of a
localic topos comes about by thinking of the underlying locale as a
complete Heyting algebra \cite{I1}.

Now the original motivation for this work was the idea of Isham that 
Physics should be done in a topos which would lie over a locale
representing all classical worlds, or perhaps classical states of an
observer. 

However, in a later paper \cite{I2}, Isham shows that the internal logic of
a topos can give us only intuitionistic logic, not quantum logic; and
that it is therefore necessary to revert to a realist interpretation
of quantum mechanics. 

This is a profound weakness in a promising new direction. Passing to
quantum topoi resolves this, since the internal logic of a quantaloid
is quantum logic. this is technically difficult to show, but no real
surprise, considering that we have a nondistributive lattice with a
noncommutative multiplication at the base.

Categorically minded readers will no doubt know that Grothendieck
topoi have two definitions, one as above i. e. sheaves over a site,
the other axiomatic. We do not know of an intrinsic characterization
of what we have called quantum topoi, but would like to have one.

\bigskip

{\bf E. Summary of properties of quantaloids}

\bigskip

Before we go on to try to create a physical theory; let us think about
how our list of desiderata above correspond to the structure of
quantaloids and quantum topoi in general.

Quantaloids have a very rich family of quotient spaces, making it easy
to filter out information. This makes us optimistic about desiderata $
\alpha , \gamma , \zeta $. We are already in a category of sheaves,
which helps us with $ \beta $ , and the algebraic structure of a
quantaloid has noncommutative multiplication and a nondistributive
lattice, dovetailing with $ \delta $ and $ \zeta $.  So a priori, it
seems that we have a useful mathematical setting. The fact that it is
a far reaching generalization of the theory of metric spaces as well
is also suggestive.

Finally, we would expect the quantum theory of gravity to resemble
quantum theory as we know it. Since each quantum mechanical system
already can be described as a quantale, we have a starting point which
is physically familiar, except for the abstract language, which the
author hopes the physical reader will eventually learn to love.

\newpage

{\bf IV. BUILDING A QUANTUM TOPOS FROM RELATIVISTIC OBSERVERS}

\bigskip

{\bf A. Physical overview}

\bigskip

Now we want to construct a quantum topos which would be an appropriate
setting for quantum gravity. 

First let us state the problem physically. We believe that an
operational interpretation of quantum mechanics means that only
positions of regions as they appear to observers, and correlations of
apparent positions, can appear in the theory. Observable regions for
each observer form a lattice. To form the kinematics of our theory we
must find some way to combine the lattices. The dynamics of the theory
then must constrain the possible correlations, and their time
evolution. A reasonable approach would be to search for a quantaloidal
nucleus to implement the dynamics. 

\bigskip

{\bf B. General Program}

\bigskip

The mathematical program is the following:

\bigskip

1. Make a quantum mechanical model of an observer in general
relativity. 

\bigskip

2. Construct the quantale corresponding to the observer. 

\bigskip

3. Construct the quantalic nucleus corresponding to the red shift
relating the observer to a region.

\bigskip

4 Form the quantaloid of all observers for the relational setting.

\bigskip

5. Study the quantum topos of sheaves over the category of
observers. As explained above, it is itself a quantaloid.

\bigskip

6. Impose Einstein's equation as a quantaloidal nucleus on  the
quantum topos.

\bigskip

Now we have not yet completed this program. We shall carry it out as
far as we know how, then give an outline of the remaining
steps. We discuss below the possibility that some further shift in
point of view will be necessary.

\bigskip

{\bf C. Observers. One eye and two.}

\bigskip

An observer at a point in a spacetime observes incoming information
about the location of past events on a 2-sphere of null lines. This
2-sphere has an action of the Lorentz group which coincides with the
action of the group of fractional linear complex transformations on
the Riemann sphere $CP^1 = S^2$. In other words, the 2-sphere of null
lines inherits a complex structure from the action of the Lorentz
group on Minkowski space.

A single observer who only marked the apparent position of an event on
a copy of $CP^1$ would not be able to observe the distance of the
event, and would not be able to infer its time either. To produce a
mathematical description of an observer who could make such a
determination, we could either combine two nearby observers with two
nearby copies of $CP^1$; or else think of the observer as nontrivially
wide, and keep track of rays from a common event which impinged on
different points of the 2-sphere at slightly different angles, which
did not converge at the center. The second possibility is perhaps more
physical, but we choose the first in order to construct a
mathematically idealised observer which would be easier to quantize.

So for us an observer has ``two eyes,'' by which we mean two nearby
copies of a $CP^1$ of null lines, with a parallelism defined between
them, from which the apparent position of a distant event could be
defined in an ideal past in Minkowski space via parallax.

We then want to have a family of such observers, with relative
positions and orientations specified, and to keep track of the
correlations between their observations of events in some past region,
in order to reproduce the geometrodynamics of some spacetime region in
their common past.
Classically, we could organize our observations as subsets of the
cartesian product of the $CP^1 s$ where correlations appeared. The
results would be somewhat complicated even classically by the presence
of gravitational lensing and consequent multiple images. regions in
event horizons would not appear unless we had observers inside them.

Now again classically, we could embed such a description in a quantaloid. 
The objects would correspond to the observers, and the hom lattices
would be the lattices of all subsets of the cartesian products of the
corresponding $CP^1 s$, a construction we referred to as the category
of relations above, over the category whose objects are the observers
and whose morphisms are functions on the spheres. Let us refer to this
quantaloid as the  (classical) Relational Observation Quantaloid..

If we wanted to include limitations on the accuracy of distance
measurements in a description of classical relativity, we could do so
by constructing a quantaloidal nucleus on the ROQ. (This is a
philosophically unattractive construction, since in classical theories
the disturbance of the system by measurements is ignored, and since
the Planck scale is a quantum effect, but it is worth thinking about
as a comprehensible toy problem).

We would then use a structure analogous to the construction of Lawvere
mentioned above to construct a quantaloid of successively fuzzy
geometries on the copies of $CP^1$ and their products.

At this point we must make an important observation. In general
relativity, the Weil tensor can induce a shear on a congruence of null
geodesics. This means that, to a two eyed observer, a point in the past
does not necessarily appear as a pair of null lines which form a plane
together with the segment connecting the centers of the two eyes. The
set of all possible apparent past events would appear as a circle
bundle over the past light cone, where the circular parameter would be
given by the dihedral angle between the two planes determined by the
two null rays. This can also be described as the cartesian product of
the two 2-spheres, with the diagonal removed, since diagonal points
would appear at infinity. We are using the assumed local parallelism
between the two 2-spheres, which can be thought of as parallel
propagation along the segment joining their focal events.

Now if we want to pass to a theory based on quantum mechanical
observers, we need to study the Hilbert space $L^2(CP^1 )
$ , together with tensor products of copies of it with itself to
represent observers. We would also need to keep track of the action of
the Lorentz group on these Hilbert spaces to compare observers in
moving frames. In short, we need the structure of the Hilbert spaces
of observers as representations of the Lorentz group; in particular in
order to model the relationships between moving observers.

This brings us directly to consider the Gelfand representations.

As we mentioned in the section on state sum models, the mathematical 
structure of the unitary representations of the
Lorentz group is a quantum geometry of Minkowski space. It is not
surprizing that it gives us a description of an ideal quantized observer in
general relativity.

In his study of the unitary representations \cite{GEL}, Gelfand studied the
Hilbert spaces of various homogeneous spaces for the Lorentz group,
and related them by means of integral transforms, to obtain the
representations and study their behavior under tensor product.

Three important homogeneous spaces for our purpose are the $CP^1$ of
null lines through an event, the past null cone of a point NC, and the
complex plane $C^2$. The last gets its action from the isomorphism of
the Lorentz group with SL(2,C). Gelfand shows that there is an
invertible integral transform between $CP^2 \otimes  CP^2 $ and $C^2$
, where the restricted product discussed above is meant. 

This means that as representations of the Lorentz group under the
natural actions

\bigskip

$L^2 (CP^2 \otimes  CP^2 )  \cong L^2(C^2 )$

\bigskip

Now the decomposition of these function spaces as irreducible
representations is known. A single $CP^2$ gives a single
representation R(0,0). It can give any nonzero $\rho $ if we modify
the action of the group to include a power of the Jacobian. The
Hilbert 
space of the past cone contains all the
Gelfand representations with zero k and arbitrary $\rho$.

\bigskip

$ L^2 (NC) \cong \int R(0, \rho) d \rho$

\bigskip

which is a direct integral of representations in the sense of Mackey
\cite{Mack}.

\bigskip

Now Gelfand's integral transform tells us that the Hilbert space on
the product of two copies of $CP^1$, the Hilbert space of a two eyed
observer is equal to the Hilbert space on $C^2$, which is the sum over
k of the integral above:

\bigskip

$ L^2 ( CP^1 \otimes CP^1 ) \cong {\sum }_k \int R(k, \rho ) d\rho $.

\bigskip

We now take this as a construction of a Hilbert space for a two eyed
observer with perfect vision. We denote it $ {H^B}_i $ for the
binocular Hilbert space of the ith observer.

The combination of Gelfand representations which appear on the cone is
precisely the subset of balanced representations which appear in the
BC model \cite{BC}. The set which appear over $C^2$ is a full tensor
category. If the BC state sum model were extended to include all those
representations, it would become a topological field theory.

The observation that a classical two eyed observer would see points as
lying in a circle bundle over spacetime translates in the quantum
version into saying that
the constraints in the BC model are partially relaxed for distant
observers. We shall use this idea below to make a conjecture as to the
dynamics of quantum gravity.

\bigskip

{\bf D. Redshifts and projections. Kinematics}

\bigskip
 
In a physical example of the relational setting, there would have to
be a projection on ${H^B}_i$ corresponding to the information which
could flow from the observed region to the observer. This is closely
related to the phenomenon of the red shift in general relativity. As
we approach the event horizon of a black hole, the red shift tends to
infinity, so the flow of information outward goes to zero. The finite
information theorems cited above are consequences of this.    

Since any observer must have a mass bounded below by the uncertainty
principle and above by its Schwartzschild mass, real observers with
two eyes would see quotient spaces of the ideal observer constructed
above. 

This part of our program has not been completed. The quotient Hilbert
spaces of mutually moving observers will not completely overlap, since
rest energies in their frames will differ.

Assuming we are able to formulate this, the space of projections on
the physical Hilbert space of each observer will form a
quantale. Extending this by the category of relations construction
will form the quantaloid of observation of any ensemble of physical
observers, whose objects will be any set of the physical observers,
to which will correspond the lattice of all correlated observed
positions observed by the set of observers, combined by taking the
tensor product of lattices.

\bigskip

{\bf DEFINITION:} {\it The physical quantum topos is the topos of sheaves
over the quantaloid of observation in the above paragraph.}

\bigskip

This is a mathematically useful way of describing all lattices of
simultaneous observed positions.

\bigskip

{\bf E. Dynamics, a conjecture}

\bigskip
So to formulate a theory, we need some way of computing the
probability that some probe will appear simultaneously in some
apparent regions to some set of real observers. We can describe probes
dually as future light cones in the past of the observed region. They
would also have projections on their dual or time reversed Hilbert
spaces, due to the theorems on finite information transfer.

The mathematical similarity between the ideal observers and the
construction of the BC model is not surprizing. A two eyed observer is
essentially observing a long thin bivector represented by a
triangle. The BC model used the bivectors of a triangulation as basic
variables \cite{BC}. We can interpret the constraint of the BC model
as saying that when the nearby observers on its vertices observe on
another the images they see are not unfocussed by a shear because the
space between them is locally flat.

This leads to the following :

\bigskip

{\bf CONJECTURE:} {\it The relational probabilities for the apparent
positions of a probe in a physical quantum topos can be calculated by
tensoring the state of a probe into the appropriate site of a BC model,
then   tensoring the resultant representations through the rest of the
triangulation, and measuring the probability amplitudes on the future
boundary. A single triangulation will suffice if it is fine enough to
contain all of the physical information which the real observers can see.}

\bigskip
Any probe will have the effect of introducing unbalanced
representations into the state sum. This will drive it toward being a
topological field theory. A sufficiently strong probe, or a
sufficiently large number of them will make the region appear
topological. The physical sign of this will be that outside observers
will see an uncorrelated thermal flux, with no information about the
interior geometry. The tqft state will therefore look exactly like a
black hole. This is consistent with the work in relativity which
models the horizon of a black hole with the CSW tqft. The relational
version of a black hole is a topological state.

This conjecture is at a very preliminary stage.  We state it despite
its vague formulation because it is at the nexus of a beautiful
combination of ideas. In particular, the convergence of the analysis
of observation in general relativity with the mathematical foundation
of the state sum models seems compelling, at least to its possibly
doting father.

\bigskip

{\bf F. Summary}

\bigskip

To say what we have proposed in nonmathematical language, it may be
helpful to compare a small region of spacetime to the appearance of a
chamber through a small hole in a thick wall at a finite
temperature. The region inside seems fuzzy, because the thermal state
of photons in equilibrium with the walls obscures the details.

We would not be tempted to think that the fuzzy apparent geometry was
real, because we could always remove the objects inside the chamber
and measure their shapes.

In the case of a small region, we cannot dissect it or enter it. The
fuzzy observed geometry is all we ever detect. If we think of virtual
processes in which the small region interacts with the exterior, the
fine details of some hypothetical unobservable interior geometry could
not be communicated to the exterior. Feynman diagrams should properly
only be integrated over the fuzzy geometry.

A mathematical description of fuzzy geometry is a difficult matter. It
is not really a finite point set. rather it is a complicated lattice
of minimal observable regions. Quantaloids are the natural language to
study such geometry.

\bigskip

{\bf G. Further directions. Not categorical enough?}

\bigskip

The foundation of the work up to this point is the notion of an
observed region. We are asking for probability amplitudes for 
where some material probe appears to be
to different observers. In making this the point of departure, we are
ignoring the specific physical character of the probe.

It is not entirely clear that we are justified in doing this. Quantum
gravity only becomes significant at energies where large neutral
probes would become unstable and likely to decompose. Highly curved
regions of spacetime would produce strong excitations of the local
matter fields; turning them off may be completely unphysical.

So we may have to refine our approach to ask for regions where an
elementary particle of a given type may appear; the excitations of
gauge fields from the standard model may mean that the correlations
for different types of probes with different charges are different.

Put differently, the quantum theory of gravity may well not exist
except as a sector of a geometric unified theory. The geometric
interpretation of gauge theory suggests as much.

Now it is interesting that still another branch of Mathematics exists,
which could provide a natural setting for such an approach. I am
referring to the theory of Grothendieck categories (same Grothendieck,
different categories). 

In this body of work, A space is represented by the tensor category of
coherent or quasicoherent sheaves over it. This category is
axiomatized, and abstract Grothendieck categories are treated
intrinsically as topological structures, with the role of regions
played by certain subcategories \cite{Ros}.

Without going into the technical details, coherent sheaves are a
generalization of bundles; so the particle physics in a region of
spacetime could be thought of as a Grothendieck category, in which the
bundle specifies the particle type, while the tensor product contains
the interactions, thought of as Feynman vertices.

The theory of Grothendieck categories has turned out to be a powerful
mathematical tool; it is the foundation of the theory of
noncommutative scheme theory, i.e. the effort to understand algebraic
geometry with noncommuting variables.

There is an important part of semiclassical general relativity which
has not fitted in a natural way into our quantum topos picture. We
refer to the Unruh and Hawking radiation effects. Both of these link
the spacetime geometry to a thermal state of the matter fields.

It is much easier to include this in a Grothendieck
categorical approach, where the matter fields are included into the
fundamental spacetime structure.

Perhaps a quantum Grothendieck topos with a thermal functor will
emerge as the final setting.  

As I have learned more about the categorical approaches to pointless
topology and geometry, it has seemed to me that there was massive
parallelism with the interesting issues in quantum Physics and
relativity. The connection between Grothendieck categories and
Feynman's approach to quantum field theory is a striking example of
this. 

\bigskip

\bigskip

{\bf APPENDIX A. Sup complete lattices, locales presheaves and topoi.}
\bigskip

{\bf DEFINITION:} {\it A partially ordered set is a set equipped with a
relation $\leq$ such that
\bigskip

1. $a\leq b$

\bigskip

2. if $a\leq b$ and $ b\leq c$ then a $\leq c$

\bigskip

3. if $a \leq b$ and $b\leq a$, then a=b.}

\bigskip.

{\bf DEFINITION:} {\it A lattice is a partially ordered set with (finite) sups and
infs. That is for any two elements a and b, and therefore for any
finite set of elements, there exist elements $a \wedge b$ and $ a\vee
b$  such that any element less than both a and b is less than $a \wedge
b$ and dually for $a \vee b$.}

\bigskip

{\bf DEFINITION:} {\it A sup complete lattice is a lattice with infinite sups.
That is, for any collection I, finite or infinite, of elements of the
lattice, there exists an element $ \vee _I $ such that any element
greater than all the members of I is also greater than $ \vee_I$}

It is a simple theorem that a sup complete lattice also has infinite
infs. 
\bigskip

{\bf DEFINITION:} {\it A frame or locale is a sup complete lattice which
satisfies the distributive law}

\bigskip

$ a \wedge (\vee _{i \in I} b_i) = \vee _{i \in I} (a \wedge b_i)$

\bigskip
 
The most obvious way to obtain a sup complete lattice is to take the
set of subsets of a set with the obvious set theoretical operations. 

Given sup-complete lattices L,M; the hom set Hom (L,M) is the set of
order preserving maps from L to M It is itself a frame or locale under
pointwise inequality. $L^{op}$ is the lattice whose
members are the same as those of L but with reversed order, and 
\bigskip

$L
\times M = Hom (L^{op}, M)$

\bigskip

just as for vector spaces. The operations of a quantaloid make use of
these operations on sup complete lattices.

The open subsets of a topological space form a frame or locale. The
category of frames has for objects frames and for morphisms maps of
lattices preserving $ \leq$ and satisfying the law

\bigskip

$ f(\vee a_i) \leq \vee f(a_i) $.

\bigskip

We call f a sup-lattice morphism

In general, sup lattice morphisms do {\bf not} respect infinite infs.

The category of locales is the opposite category to the category of
frames. In other words, a locale map is a frame map interpreted as
going in the opposite direction.

The motivation for this is largely in the example of topological
spaces. A continuous map takes open sets backwards into open sets, but
not in general forwards. Thus, locales generalize topological spaces,
and locale maps generalize continuous maps.

A locale is a special case of a quantale, in which $ \wedge $ and $
\circ$ coincide, $ \wedge $ distributes over $ \vee $ 
and $ \circ $ is commutative. 

Locales can be treated very similarly to topological spaces, but they
can be much more general. In fact, there exist pointless
locales \cite{Mac} .

It has proven very useful to think of a locale as a category whose
objects are the elements of the locale, with one morphism from a to b
if $a \leq b$ and none otherwise. This category is called the {\bf
  site} of the locale.

This definition enables us to consider structures similar to the ones
familiar in Physics, such as bundles, defined on any locale.
\bigskip

{\bf DEFINITION:} {\it A presheaf on a locale is a contravariant functor
from its site to the category of sets.}

\bigskip

This definition unpacks to a set for every element of the lattice with
a restriction map from any element to any smaller element.

The physically minded reader might find it a useful exercise to check
that if the locale is a manifold then the local sections of a bundle
form a presheaf.

A presheaf which satisfies a certain glueing property is called a
sheaf.

Now in ordinary set theoretic terms, a sheaf is a presheaf which
satisfies the following simple condition:

\bigskip

{\bf DEFINITION:} (glueing property) {\it if S is a presheaf over a space X 
such that if
{$U_i$} is a cover of U and for all $ U_i$ in the cover $ p(U_i) \in
S(U_i)$ such that

\bigskip

$  p( U_i) \mid _{U_i\cap U_j} =p( U_j) \mid _{U_i\cap U_j} $

\bigskip

then there exists a unique $p(U) \in S(U)$ 

\bigskip

with

\bigskip

$ p(U) \mid U_i = p(U_i)$ for all i, 

\bigskip

where $\mid $ denotes the
restriction map of the presheaf.} 

\bigskip

Functions and cross sections of
bundles form sheaves.

Now in the general setting of locales or categories, it has turned out
to be useful to define a presheaf as a contravariant functor, and a
sheaf as a presheaf that satisfies the glueing property, but only for
a restricted set of covers.

The sets of covers for which this definition of covers turns out to be
useful are called {\bf Grothendieck topologies}. We will not list
their axioms here since they are somewhat technical, and since they
are subsumed by quantalic nucleii on Rel(C) as discussed in III.B. 
see \cite{Mac}.
\bigskip

{\bf DEFINITION:} {\it a category with a Grothendieck topology is called a
Grothendieck site. The category of sheaves over a Grothendieck site is
called a Grothendieck topos.}

\bigskip

The category of sheaves over a locale with the Grothendieck topology
of all open covers is called a {\bf localic
  topos}. This is not the most general example of a topos, but it is
not far from it, in that a deep theorem \cite{JT} tells us that any topos
is the category of equivariant sheaves of some locale under the action
of some semigroup.

\bigskip

The mathematical interest in topoi is largely due to the fact that
they are so similar to the category of sets that is possible to do all
branches of Mathematics in a topos, where everything we know comes out
different in various ways. 

The real power of topos theory is the ability to choose between  many
different Grothendieck topologies. Since they are a special case of
quantaloidal nuclei, the theory of sheaves over quantaloids is at
least as powerful.

Let us describe an example which has interesting connections to
quantum mechanics. Recall first the fact that a vector in the Hilbert
space of a free particle is not a function, but rather an $L^2$ 
function, which is really an equivalence class of functions, and therefore
does not have a value at any point. This suggests that ordinary
quantum mechanics should admit a pointless formulation.

Now let us describe the Scott topos \cite{Scott}. Let $ \Lambda $ be the lattice of
Borel measureable subsets of some region R in $R^3$, considered as a
category with inclusions as morphisms (it is not quite a locale). 
Consider the
Gothendieck topology of all covers of any measureable subset S of R
whose union contains S up to a set of measure zero. The sheaves over
this category with glueing over these covers form the Scott topos.

In this topos, which has no points, real numbers are exactly
measurable functions. Ordinary quantum mechanics can easily be written
in this topos. The integral is added in a straightforward way.

So we see how our new language allows us to express a certain
filtering of information which is actually necessary in quantum
physics but generally taken for granted. The intuition that
exotic unmeasurable subsets of R are  unphysical can be put on a
mathematical footing.

\bigskip

{\bf APPENDIX B. Distributive and non-distributive lattices in
  Physics} 

\bigskip

{\bf DEFINITION:} {\it A world element is a proposition about a system
which is either true or false.}

\bigskip

This definition, due originally to Einstein, was one of the important
early motivations for lattice theory.

In classical mechanics, world elements correspond to Borel measurable
subsets of phase space. In quantum mechanics, on the other hand, they
are closed projections on Hilbert space. 

There is an important difference between these two types of lattice:
the first satisfies the distributive law:

\bigskip

$ a \wedge (\vee _{i \in I} b_i) = \vee _{i \in I} (a \wedge b_i)$

\bigskip

while the second does not. This was the observation which motivated
Van Neumann and Birkhoff to invent quantum logic.

The failure of the distributive law is a rather abstract way to
describe many of the phenomena that make quantum mechanics so special
and mysterious. For example, in the two slit experiment, the
probability that the particle goes through both slits and then hits
some spot is not the sum of the two probabilities of going through
each slot then hitting the spot. 

On the other hand, if we study only operations on a system in quantum
mechanics which correspond to some subset of the operators which
commute with one another, then we get a distributive lattice, and in
fact the multiplication of the operators corresponds to the
intersection of the world elements. (In the two slit experiment, the
observation of the position of the spot happens at a later time than
passage through the slits, and therefore does not commute with them).

In the quantale associated to a quantum mechanical system, sublocales
correspond to commutative subalgebras of the $C^*$ algebra.

\bigskip
{\bf ACKNOWLEDGEMENTS:} The author wishes to thank David Yetter for
helpful conversations. The Author was supported on FQXi grant
RFPI-06-02.

\bigskip

{\bf THIS IS ONLY A PRELIMINARY VERSION. THE BIBLIOGRAPHY IS
  INCOMPLETE.}


\bigskip
\begin{thebibliography}{xxxxxx}

\bibitem{GEL} I. M. Gelfand, M. I. Graev and N. Ya Vilenkin
  Generalized Functions, vol 5: Academic Press, 1966

\bibitem{Mac} S. MacLane and I Moerbeck Sheaves in Geometry and Logic
  Springer, 1992

\bibitem{JT} A. Joyal and M. Tierney An Extension of the Galois
  Theory of Grothendieck, AMS 1984

\bibitem{G} R. P. Gylys Sheaves On Quantaloids Lithuanian Mathematical
  Journal vol 40 No. 2 p 105, 2000

\bibitem{Scott} D. Scott Boolean models and nonstandard analysis p. 87
  in
  Applications of Model theory to algebra analysis and probability
  ed. W. Luxemburg, Holt Rinehart Winston 1967

\bibitem{P} A. Pitts Applications of sup-lattice enriched category
  theory to sheaf theory Proc. London Math. Soc. 57 (3) 1988 433-480\


\bibitem{L}  F. W. Lawvere Metric spaces, generalized logic and closed
  categories, Rend. Sem. Mat. e Fis. Milano 1973 135-166

\bibitem{St} J. Stachel, Einstein from B to Z Birkhauser 2002

\bibitem{BC} J. Barrett and L. Crane A Lorentzian signature model for
  quantum general relativity CQG 17 (2000) p3101-3118

\bibitem{C1} L. Crane, Categorical Geometry and the Mathematical
  Foundations 
of Quantum General Relativity; Contribution to the Cambridge
University Press 
volume on Quantum Gravity, D. Oriti ed. 

\bibitem{TV} 	 `t Hooft, G. ; Veltman, M. ; Speiser, D, Diagrammar
  Plenum Publishing Corp 1974.



\bibitem{I1} C. J. Isham
 Topos Theory and Consistent Histories: The Internal Logic of the Set of all Consistent Sets
Int.J.Theor.Phys. 36 (1997) 785-814

\bibitem{I2} A. Doering, C.J. Isham
A Topos Foundation for Theories of Physics: I. Formal Languages for Physics;
quant-ph/0703060

\bibitem{Ros}Alexander L. Rosenberg, Noncommutative schemes Compositio Mathematica
Volume 112, Number 1 / May, 1998

\bibitem{Mack} G. Mackey,The theory of unitary group representations,
University of Chicago Press, 1976


\end{thebibliography}
\end{document}